\documentclass[prd,twocolumn]{revtex4}
\usepackage{graphicx, epsfig}
\usepackage{color}
\usepackage{mathrsfs}
\usepackage{amsmath}

\newcommand{\be}{\begin{equation}}
\newcommand{\ee}{\end{equation}}
\newcommand{\bea}{\begin{eqnarray}}
\newcommand{\eea}{\end{eqnarray}}

\newcommand{\gapp}{\mathrel{\raise.3ex\hbox{$>$}\mkern-14mu
              \lower0.6ex\hbox{$\sim$}}}
\newcommand{\lapp}{\mathrel{\raise.3ex\hbox{$<$}\mkern-14mu
              \lower0.6ex\hbox{$\sim$}}}

\begin{document}
\title{Analytic explanation of the strong spin-dependent amplification in Hawking radiation from rotating black holes}
\author{De-Chang Dai and Dejan Stojkovic}
\affiliation{HEPCOS, Department of Physics,
SUNY at Buffalo, Buffalo, NY 14260-1500}


\begin{abstract}

\widetext
Numerical studies of black hole greybody factors indicate that Hawking emission from a highly rotating black hole
is strongly spin dependent, with particles of highest spin (gravitons) dominating the energy spectrum. So far, there has been no analytic explanation or description of this effect. Using "gravitomagnetism", or the formal analogy between the Maxwell's field equations for electromagnetism and Einstein's equations for gravity, we were able to establish a link between the spin of the rotating black hole and spin of an emitted particle. Namely, the intrinsic spin of the particle creates a ``mass dipole moment" which interacts with external gravitomagnetic field whose source is the rotation of the black hole. We showed that a rotating black hole prefers to shed its spin, i.e. tends to emit particles with the spin parallel to its own. We also showed that the probability for emission grows with the increasing spin of the emitted particles. The amplification factors can be huge if a black hole is highly rotating, i.e. close to extremal. When applied to central galactic black holes, the same physical mechanism indicate that particles orbiting around these black holes should have spins strongly correlated with the spin of the black hole, which may have implications for cosmic rays believed to be coming from these regions of space.
\end{abstract}


\maketitle

\section{Introduction}
One of the first use of the term "superradiance" was by R.H. Dicke in 1954 \cite{dicke} to describe an effect in which disordered energy of various kinds is converted into coherent energy. In the context of classical black hole physic, superradiance is a classical phenomenon in which an amplitude of an outgoing wave after the reflection is greater than the amplitude of the ingoing wave \cite{Zeldovich} .
This situation can happen in the background of a rotating black hole \cite{unruh}. Such a background contains an ergosphere (the region between the infinite redshift surface and the event horizon)
which allows an incident wave to take away some of the rotational energy of the black hole and get amplified after reflection, giving a negative absorption coefficient (i.e. the reflection coefficient greater than one).

\begin{figure}
	\epsfig{file=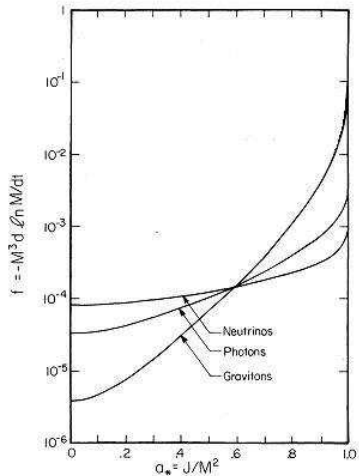,height=70mm}
	\caption
	{\label{dp} Hawking radiation of particles with different spin from a rotating black hole. A highly rotating black hole ($a_* \approx 1$) prefers to emit particles of higher spin. While Hawking radiation from a non-rotating black hole ($a_* \approx 0$) is dominated by emission of lower spin particles, the probability for emission of a graviton by a highly rotating black hole is hundred times greater than the probability of emission of a neutrino and ten times greater than the probability of emission of a photon. This implies that the amplification is strongly spin dependent, favoring higher spin particles. [Data taken from \cite{Page:1976ki}.]}
\end{figure}

Soon after the discovery of quantum Hawking radiation from a black hole, it was noticed that spontaneous emission from a black hole can also get amplified taking away rotational energy of the black hole.
There are numerous numerical studies of this effect, but the first calculation was done by Don Page \cite{Page:1976ki}. It appears that this amplification is very much spin dependent, with emission of higher spin particles strongly favored. As calculated by Don Page \cite{Page:1976ki}, the probability of emission of a graviton by a highly rotating black hole is hundred times greater
than the probability of emission of a neutrino and ten times greater than the probability of emission of a photon (see Figure~\ref{dp})\footnote{For related work in higher dimensions and string theory see \cite{Frolov:2002xf,Frolov:2002as,Dai:2007ki,Dai:2006hf,Casals:2008pq,Jung:2005nf,Bredberg:2009pv}.}.
However, so far there is no analytic theoretical framework which describes and explains the physics of this amplification.

While the spin dependent amplification of Hawking radiation is also often called superradiance in the spirit of Dicke's definition in \cite{dicke}, it is somewhat different from the spin dependent effects in \cite{Zeldovich} or \cite{unruh} which crucially depend on the negative absorption coefficient. For example the superradiant amplification as defined in \cite{Zeldovich} or \cite{unruh} is not possible for fermions \cite{unruh,Frolov_book}. If the incident wave is made of fermions, then the reflected wave can not get amplified due to Pauli exclusion principle, since all the available states are already occupied. However, the process of Hawking radiation is much more complex and fermions can benefit from the rotational energy of the black hole without having the negative absorption coefficient, as the plot in Figure~\ref{dp} shows (higher values of the rotating parameter $a_*$ imply higher probability for emission of a neutrino).

In this paper, we will try to explain the effect of the strong spin-dependent amplification in Hawking radiation from rotating black holes using the formalism of gravitomagnetism. In order to predict what an asymptotic observer would measure in the case of the Hawking radiation, one has to understand what is going on near the horizon where particles are created, and near the gravitational potential barrier which created particles have to penetrate.
Indeed, the spectrum of Hawking radiation, i.e. energy $dE$ emitted by a rotating black hole per unit time $dt$ and per unit frequency $d\omega$ is
\be \label{shr}
 \frac{d^{2}E}{dtd\omega
}=\sum_{l,m}\frac{\omega }{e^{(\omega -m\Omega)/T_{h}}-(-1)^{2s}} \frac{N_{l,m}|A_{l,m}|^{2}}{2\pi} \, ,
\ee
where $T_{h}$ is the Hawking temperature, $l$ and $m$ are the total angular momentum quantum numbers, $s$ is the spin of the particle, $\Omega$ is the angular velocity of the black hole,  $N_{l,m}$ is the number of available degrees of freedom and $A_{l,m}$ is the absorption coefficient. The absorption coefficient $A_{l,m}$ actually determines the transmission cross section of a particle interacting with the black hole potential barrier, i.e. the probability that a created particle will penetrate the barrier. We can schematically represent Eq.~(\ref{shr}) as a product of two terms
\be \label{bbgb}
 \frac{d^{2}E}{dtd\omega
}=BB \times GB \, ,
\ee
where $BB$ stands for the thermal black body term, while $GB$ stands for the greybody term. The black body term gives the probability that a certain particle is thermally produced near horizon, while the greybody term modifies the thermal radiation due to the existence of the potential barrier which the created particle has to penetrate. The absorption coefficient $A_{l,m}$ which crucially depends on the potential barrier directly gives the greybody factor.
The greybody factor is obtained by solving the radial equation of the particle propagating in the background of the rotating black hole with mass $M$ and angular momentum $J$ (see e.g. \cite{Frolov_book} Sec. 4.8.1)
\begin{eqnarray} \label{req}
&&\Delta^{-s} \frac{d}{dr} \left( \Delta^{s+1}\frac{dR}{dr} \right) \nonumber \\
&&+ \left[ \frac{K^2-2is(r-M)K}{\Delta} +4is\omega r -\lambda  \right]R=0
\end{eqnarray}
where $s$ is the spin of the particle, $a=J/M$ is the rotational parameter, $\Delta \equiv r^2-2Mr+a^2$, $K \equiv (r^2+a^2)\omega -am$, and the eigenvalue $\lambda \equiv E- s(s+1) +a^2\omega^2 -2am\omega$ with $E$ being the total energy of the particle.

The terms $m\Omega$ in the thermal part of Eq.~(\ref{shr}) and $am$ in the eigenvalue $\lambda$ in Eq.~(\ref{req}) indicate that spin-spin interaction of the black hole and created particle may play an important role in Hawking radiation from a rotating black hole.
Not that the quantum numbers $l$ and $m$ refer to the total angular momentum of the particle which in general includes spin, while in the scalar case they refer to the orbital angular momentum). Due to the highly non-linear and complicated nature of gravity, it is not easy to see the exact nature of this spin-spin interaction and the mechanism through which it works.  However, using the formalism of gravitomagnetism, it is possible to write down the (approximative) analytic spin-spin interaction, which may shed more light on the strong spin-dependent amplification in Hawking radiation from rotating black holes.

\section{The spin-spin interaction between the black hole and particle}

In what follows, we will demonstrate that, due to spin-spin interaction of the black hole and created particle, particles created with spin parallel to that of a black hole will minimize their total energy. The higher the spin of the particle is, the stronger spin-spin interaction is, which lowers the energy of the particle.
Near the horizon, particles are created thermally, i.e. all the available degrees of freedom whose total energy is less or equal to the temperature of the black hole will be created. This implies that among particles of the same sort, mostly those with spin parallel to that of a black hole will be created taking away black hole angular momentum (spin). Among massless (or almost massless) particles of the same frequency, mostly those of the higher spin will be created since their total energy is lower. The spin-spin interaction also lowers the potential barrier that the emitted particles have to tunnel through.  Thus, these two effects work synergetically to amplify emission of particles of higher spin from a rotating black hole, as it was observed in numerical studies.

We will use the formalism of "gravitomagnetism", which  is a formal analogy between the Maxwell's field equations for electromagnetism and Einstein's equations for gravity, in an approximation valid far from isolated sources, and for slowly moving test particles. While this approximation appears restrictive, it may give us the right order of magnitude estimates.
The metric of the rotating object of mass $M$ and angular momentum $\vec{J}$, in the weak field approximation, can be written as
\begin{equation}
\label{geo}
ds^2=-(1-2\Phi )dt^2-4\vec{A}\cdot d\vec{x} dt+(1+2\Phi )\delta_{ij} dx^i dx^j
\end{equation}
where $\vec{x} = (x,y,z)$, $r^2=x^2+y^2+z^2$, $\Phi \sim M/r$ and $\vec{A}\sim \vec{J} \times \vec{x}/r^3$.
In gravitoelectromagnetism \cite{Mashhoon:2000he} (GEM), this gravitational field can be expressed in terms of ``electric" and ``magnetic" fields.

\begin{eqnarray}
\vec{E}&=&- \nabla \Phi -\frac{\partial}{\partial t}(\frac{\vec{A}}{2})\\
\label{B-field}
\vec{B}&=&\nabla \times \vec{A}   \label{b}
\end{eqnarray}

Motion of a test particle of mass $m$, in a given metric whose line element is $ds$, follows from the Lagrangian ${\cal L}=-mds/dt$. From Eq.~(\ref{geo}) we get
\begin{equation}
{\cal L}=-m(1-v^2)^{1/2}+m \gamma (1+v^2)\Phi -2m \gamma \vec{v}\cdot \vec{A}
\end{equation}
Here, $\gamma$ is the Lorentz factor and $\vec{v}$ is the velocity of the test particle. If $\partial \vec{A}/ \partial t =0$, the analogous Lorentz force can be written as
\begin{equation}
F=q_E \vec{E}+q_B \vec{v} \times \vec{B}
\end{equation}
with $q_E=-m$ and $q_B=-2m$.

\begin{figure}[h]
   \centering
\includegraphics[width=2.5in]{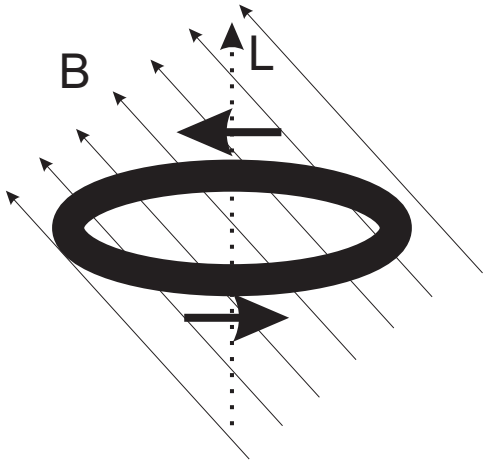}
\caption{A rotating ring with angular momentum $\vec{L}$ in a constant gravitomagnetic field $\vec{B}$.}
    \label{ring}
\end{figure}

Consider now a test particle rotating around an object described by the metric (\ref{geo}). We can approximate such a test particle with a massive rotating ring placed in a
constant gravitomagnetic field $\vec{B}$, as shown in Figure~\ref{ring}. This rotating ring will create an ``angular mass dipole moment" with
\begin{equation} \label{mdmL}
\vec{\mu}_g =q_B \frac{\vec{L}}{2m}
\end{equation}
The intrinsic spin of the particle will also create a mass dipole moment with
\begin{equation} \label{mdm}
\vec{\mu}_g =q_B \frac{\vec{S}}{2m}
\end{equation}
where $\vec{S}$ is the spin of the particle. Since $q_B=-2m$, the mass of the test particle disappears from Eq.~(\ref{mdm}) (but it does not disappear from Eq.~(\ref{mdmL}) since $\vec{L} = \vec{r} \times m \vec{v}$). This is expected since spin is an intrinsic property of a particle independent of its mass.

The torque acting on a particle due to the external gravitomagnetic field $\vec{B}$ is
\begin{equation}\label{torque}
\vec{\tau} =-\vec{\mu}_g \times \vec{B} .
\end{equation}
This torque $\vec{\tau}$ will affect orientation of the vector $\vec{S}$. [We note here that a similar result can be derived from the Post-Post-Newtonian approximation (PPN) or using Papapetrou's equations of motion for spin \cite{Schiff:1960gi,Mashhoon:1971nm,Wald:1972sz}, however, gravitomagnetism reveals a clear analogy with electromagnetism.]
 Using the formal analogy with electromagnetism,  we write down the interaction  between the mass dipole moment $\mu_g$ from Eq.~(\ref{mdm}) and an external gravitomagnetic field $\vec{B}$ from Eq.~(\ref{b}) as
\begin{equation}\label{int}
H_{int}=-\vec{\mu}_g \cdot \vec{B}=\vec{S}\cdot \vec{B}.
\end{equation}
As the energy of interaction tends to be minimized, the spin of the test particle will tend to align with the opposite direction of $\vec{B}$.

For classical evolution, the torque will force the spin of the test particle to precesses around the $\vec{B}$. The effects of dissipation will eventually align the spin with $-\vec{B}$.
For a planet rotating around a star, this torque and dissipation are small, and the planet will precess around the vector $\vec{B}$ under some small angle for long time. However, for quantum effects dissipation is not needed. For example, in the case of an elementary particle,  the interaction with $\vec{B}$ can easily change orientation of the particle's spin. For a particle, the characteristic time for flipping can be estimated from the uncertainty principle as
\begin{equation}
\label{flip}
\Delta t \sim \frac{1}{\Delta H_{int}}\sim \frac{1}{\Delta \vec{S}\cdot \vec{B}}
\end{equation}
As expected, in a stronger gravitomagnetic field $\vec{B}$ a particle flips its spin faster. The other quantum effects of interest would be spontaneous pair creation in vicinity of a rotating black hole and tunneling through the potential barrier which we will discuss later.

Consider now a black hole whose angular momentum points is $z$-direction, i.e. $\vec{J} \sim \hat{e}_z$. The vector $\vec{B}$ is then
\begin{equation}\label{B}
\vec{B}=J (\frac{2}{r^3}-3\frac{x^2+y^2}{r^5})\hat{e}_z+J\frac{3\,z}{r^5}(x\hat{e}_x+y\hat{e}_y)
\end{equation}
where $\vec{r}$ is the position vector with $r^2=x^2+y^2+z^2$.
Since it is very unlikely that a random particle with the location $\vec{r}$ will have its spin parallel to $\vec{r}$, for simplicity we remove the component of $\vec{B}$ that is parallel to $\vec{r}$. The remaining component is
\begin{equation} \label{bperp}
\vec{B}_{\perp}=-J\frac{x^2+y^2}{r^5}\hat{e}_z+J\frac{z}{r^5}(x\hat{e}_x+y\hat{e}_y)
\end{equation}
If the orbit of the particle is symmetric with respect to the origin, the components in $\hat{e}_x$ and $\hat{e}_y$ directions will cancel out each other. The net effect is then that  the spin of the particle tends to align with the opposite direction of $\vec{B}$ (i.e. to align with the positive direction of $\vec{e}_z$) to minimize its energy, which in this case is the direction of the angular momentum of the black hole. Thus, the spin of the particle tends to align with the direction of the angular momentum of the black hole.

We can estimate some characteristic times for this process to occur. Consider first a test particle near the Sun's equator. $M_\odot=1.9891\times 10^{30} kg$, $R_\odot = 1.392\times 10^9 m$, and $J_\odot \sim 5\times 10^{41}kg \ m^2/s$, and $\Delta S\sim \hbar$. From equation (\ref{flip}), the flipping time is about $7.3\times 10^{12}s$, which is very unlikely to be observed. However, if we replace the Sun with an extremal black hole of mass $1M_\odot$, the flipping time becomes $4.8\times 10^{-6}s$. Therefore, we can expect the spins of particles around extremal (or close to extremal) rotating black holes to be practically all aligned with the angular momentum of the black hole (in the absence of external disturbances). This polarization of spins in the background of highly rotating central galactic black holes may perhaps be observed in future.
Active Galactic Nuclei (AGN), which are believed to harbor large highly rotating black holes, appear to be the source of radiation of many different kinds of particles. Detecting the correlation between the spin of the black hole and that of the emitted particle (perhaps a neutral particle in order to avoid the effects of magnetic fields) would be a strong support for our findings.

\section{Consequences for Hawking radiation}

We now turn to Hawking radiation, which is a quantum effect where particles are spontaneously created in the gravitational field of the black hole.
This energy difference between different orientations of the spins of created particles may significantly affect the energy distribution of Hawking radiation. We estimate
the energy distribution of created particles from the Boltzmann distribution
\begin{equation}
P(E)=e^{-E/T}
\end{equation}
where $P(E)$ is the probability that a particle in the state with the total energy $E$ is created, while $T$ is the temperature of the background (i.e. black hole Hawking temperature). Suppose two particles are created at the same position, with different spins but with the same linear momenta. The ratio of probabilities for these two particles is
\begin{equation}\label{pratio}
P_{\rm ratio}=e^{-\Delta H_{int}/T}=e^{-\Delta \vec{S}\cdot \vec{B}/T} \, ,
\end{equation}
where $\Delta \vec{S}$ is the difference in spins of the emitted particles whose probability ratio we are calculating.
Since we saw that the particle spin $\vec{S}$ tends to be aligned with the spin of the black hole, i.e. to point to the direction opposite to $\vec{B}$, then the higher spin particle has higher probability to appear than the lower spin one. Thus the amplification is stronger for particles of higher spin.

In the context of Hawking radiation,  particles are created thermally near the horizon, i.e. all the available degrees of freedom lighter than the temperature of the black hole are created. Since the interaction term in Eq.~(\ref{int}) lowers the total energy of a particle that feels the gravitomagnetic field $\vec{B}$, this implies that among massless (or almost massless) particles mostly those of the higher spin will be created since their total energy is lower.
This is in agreement with Eq.~(\ref{pratio}). It is true that the approximation that we used is not strictly valid exactly at the horizon, but certainly captures non-trivial information about the metric of the rotating black hole, as can be seen from Eq.~(\ref{geo}).

Once a particle is created near the horizon, it has to penetrate the potential barrier in order to reach an asymptotic observer.
The same interaction term in Eq.~(\ref{int}) now lowers the potential barrier that the created particles have to tunnel through. Since most of the created particles are those of higher spins, and have spins already aligned with that of a black hole, they do not have much to compete against. Thus, these two effects work synergetically to amplify emission of particles of higher spin from a rotating black hole, as it was observed in numerical studies.

\begin{figure}[h]
   \centering
\includegraphics[width=3.0in]{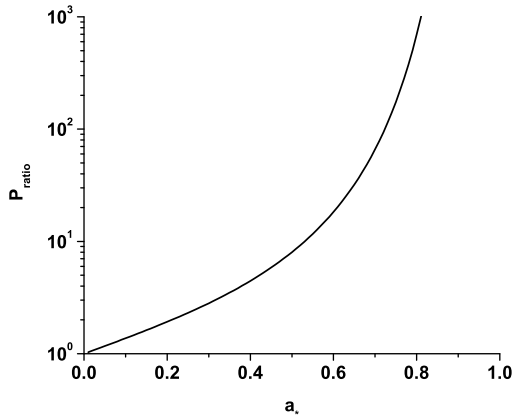}
\caption{ $P_{\rm ratio}$ (the probability ratio of emission of two particles with different spins) as a function of the black hole spin parameter $a_{*}=J/M^2$. The emitted particles spin difference $\Delta S$ is set to be $\hbar$. $P_{\rm ratio}$ grows with $a_{*}$, and the enhancement factor can easily reach several orders of magnitude, which is in qualitative agreement with numerical results in Fig.~\ref{dp}.}
    \label{ratio}
\end{figure}

Another important observation is the magnitude of the effect for highly rotating black holes. The Hawking temperature of a rotating black hole is
\begin{equation}
T_{bh}=\frac{2}{1+1/\sqrt{1 - a_*^2}}\frac{1}{8\pi M}
\end{equation}
where $a_*=J/M^2$. The temperature is lower for the higher spin of the black hole. A near extremal black hole (i.e. $a_* \sim 1$) has the temperature near zero. This significantly enhances the probability to emit a higher spin particle.  A near extremal black hole will essentially emit only highest spin particles, i.e. gravitons. To illustrate this we plot Eq.~(\ref{pratio}) as a function of the black hole spin parameter $a_*$ in Fig.~\ref{ratio}. We fix $\Delta S = \hbar$, which will give the ratio between, say, gravitons and photons, and we assume that particles are emitted near the equator of the black hole (where the expression for $B$ in Eq.~(\ref{B}) simplifies with $x^2+y^2 =r^2 =R^2_{bh}$ and $z=0$). Clearly, $P_{\rm ratio}$ grows with $a_{*}$, and the enhancement factor can easily reach several orders of magnitude, which is in qualitative agreement with numerical results in Fig.~\ref{dp}. We note that the precise quantitative agreement can not be expected.
The plot in Fig.~\ref{ratio} diverges at $a_{*}=1$, but at that point the temperature of the black hole becomes zero and is therefore unattainable according to the third law of thermodynamics. The effects of backreaction will prevent this from happening. The importance of the plot in Fig.~\ref{ratio} is to show that huge amplification is possible. Indeed, the analytic expressions that we used are not correct exactly at the horizon. Therefore the magnitude of the gravitomagnetic field in Eq.~(\ref{bperp}) is only given approximatively. For example, a multiplicative factor of the order $2-3$ in Eq.~(\ref{bperp}) would correspond in one order of magnitude change in $P_{\rm ratio}$. Also, inclusion of a small $z$-dependent component in Eq.~(\ref{bperp}) (due to imperfect cancelation) will reduce the magnitude of the gravitomagnetic field bringing the results closer to the exact numerical values in Fig.~\ref{dp}.

A non-rotating black hole prefers to emit particles of lower spin (the spin practically acts as a damping factor in the effective potential of the non-rotating black hole), which is the point $a_{*}=0$ on the plot in Fig.~\ref{dp}.
For non-zero values of $a_{*}$, the spin-spin interaction between the black hole and emitted particle kicks in, which gives rise to the spin dependent amplification. As we showed, this amplification is stronger for the higher spin fields, and easily reaches several orders of magnitude in difference. This qualitatively explains the numerical results in Fig.~\ref{dp}.

Note that the fermions also benefit from the spin dependent amplification. This is in contrast with the superradiance effect as defined in \cite{Zeldovich} or \cite{unruh} that crucially depends on the negative absorption coefficient. The negative absorption coefficient makes the $GB$ term in Eq.~(\ref{bbgb}) negative, which in turn cancels the negative sign of the $BB$ term in Eq.~(\ref{bbgb}) for modes satisfying  $0 \leq \omega \leq m\Omega$. The absorption coefficient can never be negative for fermions, which implies that fermions can not make use of the superradiant modes. However the spin-spin interaction that we describe implies that the thermal black body term $BB$ in Eq.~(\ref{bbgb}) allows for the spin dependent amplification (as in Eq.~(\ref{pratio})), and is the main source of amplification for fermions.

For completeness we mention that higher $l$ modes of the scalar particles (i.e. $s=0$) can also be amplified. Scalar particles have the intrinsic angular momentum (i.e. spin) zero, but their orbital angular momentum can be non-zero, which is described by $l>0$ modes. Note that our formalism still applies, one just needs to replace the magnitude of spin with the magnitude of the orbital angular momentum, since they are both just angular momenta.

\section{Conclusions}

In conclusion, we tried to explain the effect of the strong spin-dependent amplification in Hawking radiation from rotating black holes using the formalism of gravitomagnetism, i.e. the formal analogy between the equations of electromagnetism and gravity.
The analogy is limited to weak fields and slow motion of test particles, but it may give us the correct qualitative behavior. Slow motion of test particles implies that this formalism has some limitations when applied to massless particles. Note however that the mass of the particle cancels out in Eq.~(\ref{mdm}) and the results do not explicitly depend on the particle mass. We thus believe that our conclusions are valid for both massive and massless particles.

We established a possible link between the spin of the rotating black hole and spin of an emitted particle, for which the interaction between the gravitomagnetic field and the spin of the test particle is responsible. While Eq.~(\ref{torque}) was earlier derived in \cite{Wald:1972sz} in the PPN approximation, our derivation in the context of gravitomagnetism reveals a clear analogy with electromagnetism. We then used this analogy to write down the energy of the spin-spin interaction between the black hole and emitted particle. In the context of spontaneous emission of particles, we showed that a rotating black hole prefers to shed its spin, i.e. tends to emit particles with the spin parallel to its own in order to minimize the interaction energy.
We showed that the probability for emission grows with the increasing spin of the emitted particles, in agreement with numerical studies. The amplification factors can be huge if a black hole is highly rotating, i.e. close to extremal. Therefore, a highly rotating black hole prefers to emit mostly gravitons, while emission of particles of lower spin is highly suppressed.

Since we worked in the weak field approximation, some of our results are applicable to other rotating systems. For example, a test particle rotating around the Sun will tend to align its spin with that of the Sun, albeit on very large time scales. The characteristic time for flipping the spin gets must shorter in vicinity of the black hole. When applied to central galactic balk holes, our results indicate that particles orbiting around these black holes should have spins strongly correlated with the spin of the black hole. This polarization of spins in the background of highly rotating central galactic black holes may perhaps be observed in future.

\begin{acknowledgments}
We are especially indebted to Valeri Frolov who originated the idea that the effect of superradiance could be analytically explained using the formalism of gravitomagnetism.
 D.S. acknowledges the financial support from NSF, grant number PHY-0914893.

\end{acknowledgments}

\end{document}